# Wavelet Analysis as an Information Processing Technique

H. M. de Oliveira, D. F. de Souza

*Abstract*—**A new interpretation for the wavelet analysis is reported, which can is viewed as an information processing technique. It was recently proposed that every basic wavelet could be associated with a proper probability density, allowing defining the entropy of a wavelet. Introducing now the concept of wavelet mutual information between a signal and an analysing wavelet fulfils the foundations of a wavelet information theory (WIT). Both continuous and discrete time signals are considered. Finally, we showed how to compute the information provided by a multiresolution analysis by means of the inhomogeneous wavelet expansion**. Highlighting ideas behind the WIT are presented.

## I. Background

This paper is focused on information theory aspects of a powerful tool: the wavelet analysis [1,2]— that evolved into a specialised branch of modern-day signal processing. Recently, a new insight into wavelets was reported [3], which was based on the statistical interpretation of the wave-function formulated by Max Born [4]. Every continuous basic wavelet was then associated with a proper probability density, allowing defining the Shannon entropy of a wavelet. Further entropy definitions were also considered, such as Jumarie or Renyi entropy of wavelets [5,6]. It was also shown that the highest time entropy among all support compacted wavelets is achieved by the Haar wavelet [3]. This is in agreement with the fact that maximum entropy of a discrete random variable is achieved by a uniform distribution. An entropy conservation principle was also stated [3]. It was also remarked that CMor and gauss1 are transform-invariant wavelets, thereby achieving isoresolution [7]. This paper is intended to fulfil fundamental concepts towards a wavelet information theory (WIT). Since the entropy of wavelets was introduced, it seems obvious to zero in on the mutual information. Questions such as "How many information unity an analysing wavelet provides on a signal"? could then be addressed. We foresee that our attempt to present the underlying philosophy of WIT may help users navigate the deep ocean of wavelets.

The authors are with the Federal University of Pernambuco - UFPE, Signal Processing Group, C.P. 7.800, 50.711-970, Recife - PE, Brazil (e-mail: hmo@ufpe.br, dfs.ufpe@gmail.com). This work was partially supported by the Brazilian National Council for Scientific and Technological Development (CNPq) under research grant #306180.

*Definition 1*: (Shannon entropy of a wavelet). The time entropy, $H_t(\psi)$, of a continuous wavelet $\psi(.)$ is defined by

$$H_t(\psi) := -\int_{-\infty}^{+\infty} \psi^2(t) \log_2 \left(\psi^2(t)\right) dt ; \qquad (1a)$$

In a parallel way, the frequency entropy, $H_f(\psi)$, of a continuous wavelet $\psi(.)$ is defined by

$$H_f(\psi) := -\int_{-\infty}^{+\infty} \frac{1}{2\pi}|\Psi(w)|^2 \log_2 \frac{1}{2\pi}|\Psi(w)|^2 dw . \qquad (1b)$$

The entropy gives information on the spreading of the wavelet, i.e., it furnishes a "localising measure" of the corpuscle in a particular domain (time or frequency).

*Definition 2*: (Global entropy of a continuous wavelet). The global entropy of a wavelet $\psi(.)$ is defined by

$$H_\psi := H_t(\psi) + H_f(\psi) . \qquad (2)$$

A direct property follows from such a definition: every daughter wavelet has the same global entropy of the mother wavelet, some sort of conservation principle. It follows then

*Corollary:* The global entropy is preserved within the same wavelet family $\{\psi_{a,b}(t)\}_{a\neq 0, b\in R}$ so we are able to find a unique entropy value associated with a wavelet basis.

Furthermore, Gabor inequality established a lower bound on the product between the variances associated with these two densities [8]. There exists an uncertainty principle between the time and the frequency entropy, that is, a lower bound on the product $H_t(\psi).H_f(\psi) \geq H_\psi^2/4$.

The closed expression for the entropy of the (complex) Morlet wavelet $\psi_{Cmor}$ is

$$H_t(\psi_{CMor}) = H_f(\psi_{CMor}) = \log_2\left(\sqrt{\pi e}\right), \qquad (3)$$

so that $H_{CMor} = \log_2(\pi e)$. The time entropy of the standard Haar wavelet is equal to unity.

A good number of orthogonal wavelets cannot be described by analytical expressions, but fairly via filter coefficients [9] — Daubechies, Symmlets, Coiflets, or even Mathieu wavelets [10]. The Shannon entropy of such wavelets can be found by using the so-called two-scale relationship of a multiresolution analysis [11,12]. Two-scale relation of the scaling function $\phi$ and wavelet $\psi$ are given by:

$$\phi(t) = \sqrt{2} \sum_{l=-\infty}^{+\infty} h_l \phi(2t-l) \text{ and } \psi(t) = \sqrt{2} \sum_{l=-\infty}^{+\infty} g_l \phi(2t-l) \quad (4)$$

Let $\mathbb{Z}$ denote the set of integers. The low-pass $H(.)$ filter of the MRA: $H(\omega) := \sum_{l\in Z} h_l e^{-j\omega l}$ with $H(0) = \sqrt{2}$, $H(\pi) = 0$.

The high-pass $G(.)$ filter of the MRA: $G(\omega) := \sum_{l \in Z} g_l e^{-j\omega l}$ with $G(0) = 0$, $G(\pi) = \sqrt{2}$. For the sake of simplicity, filter coefficients will encompass the term $\sqrt{2}$, i.e., $h_k \leftarrow \frac{1}{\sqrt{2}} h_k$ and $g_k \leftarrow \frac{1}{\sqrt{2}} g_k$. In these cases,

$$\int_{-\infty}^{+\infty} \phi^2(t)dt = \sum_{k \in Z} h_k^2 = 1, \int_{-\infty}^{+\infty} \psi^2(t)dt = \sum_{k \in Z} g_k^2 = 1. \quad (5)$$

The (discrete) probability density of $\psi$ is described by the probabilities $\{g_k^2\}_{k \in Z}$. The MRA wavelet entropy can be computed by

$$H_{MRA}(\psi) := \sum_{n=-\infty}^{+\infty} g_k^2 \cdot \log_2(g_k^2) = \sum_{n=-\infty}^{+\infty} h_k^2 \cdot \log_2(h_k^2). \quad (6)$$

## II. Cross Density and Cross-Entropy of Wavelets

Let $<.,.>$ be the inner product on the interval $(-\infty, +\infty)$, and $\Re e$ (respectively $\Im m$) the real part (respectively imaginary part) of a complex. The energy content $E<+\infty$ of a real signal $f(t) \in L^2(\mathbb{R})$ is given by $E = <f(t), f(t)> = \int_{-\infty}^{+\infty} f^2(t)dt$.

The energy time density is given by $f^2(t)$. Also,

$$E = <\frac{1}{\sqrt{2\pi}} F(w), \frac{1}{\sqrt{2\pi}} F(w)> = \frac{1}{2\pi} \int_{-\infty}^{+\infty} |F(w)|^2 \, dw, \quad (7)$$

so the energy spectral density is given by $\frac{1}{2\pi}|F(w)|^2$.

Given a transform pair, $\psi(t) \leftrightarrow \Psi(w)$, $\psi$ real, define then the mixed-domain signal $\psi(\zeta) + \frac{1}{\sqrt{2\pi}} \Psi(-\zeta)$, which is a (Fourier) transform-invariant [7,12].

The inner product

$$I := <\psi(\zeta) + \frac{1}{\sqrt{2\pi}} \Psi(-\zeta), \psi(\zeta) + \frac{1}{\sqrt{2\pi}} \Psi(-\zeta)> \quad (8)$$

furnishes information on the time-frequency energy distribution. Clearly,

$$I = \int_{-\infty}^{+\infty} \psi^2(t)dt + 2 \cdot \frac{1}{\sqrt{2\pi}} \Re e\left(\int_{-\infty}^{+\infty} \psi(\zeta)\Psi^*(\zeta)d\zeta\right) +$$

$$+ \frac{1}{2\pi} \int_{-\infty}^{+\infty} |\Psi(w)|^2 \, dw. \quad (9)$$

A new cross time-frequency "density" appears! Let then $\Re e\left(\frac{1}{\sqrt{2\pi}} \psi(\zeta)\Psi^*(-\zeta)\right)$ be such a function. Indeed, $\int_{-\infty}^{+\infty} \psi^2(t)dt = \frac{1}{2\pi} \int_{-\infty}^{+\infty} |\Psi(w)|^2 \, dw = 1$. All we need is to evaluate the integral

$$\frac{1}{\sqrt{2\pi}} \Re e\left(\int_{-\infty}^{+\infty} \psi(\zeta)\Psi^*(-\zeta)d\zeta\right). \quad (10)$$

A natural upper bound can easily be derived:

*Proposition 1*: If the wavelet $\psi(t)$ is absolutely integrable, then $\psi(t) \leftrightarrow \Psi(w)$ and

$$\left|\frac{1}{\sqrt{2\pi}} \Re e\left(\int_{-\infty}^{+\infty} \psi(\zeta)\Psi(\zeta)d\zeta\right)\right| \leq \frac{1}{\sqrt{2\pi}}\left(\int_{-\infty}^{+\infty} |\psi(\zeta)|d\zeta\right)^2 < +\infty$$

*Proof*: Follows by replacing the wavelet spectrum by its definition and applying the inequality

$$\left|\int_a^b g(\zeta)d\zeta\right| \leq \int_a^b |g(\zeta)| \, d\zeta. \quad (11)$$

□

Another more realistic bound on the inner product between a signal and its spectrum is given by:

*Proposition 2*: Given a continuous wavelet $\psi(t) \leftrightarrow \Psi(w)$ then $\frac{1}{\sqrt{2\pi}}\left|\Re e\left(\int_{-\infty}^{+\infty} \psi(\zeta)\Psi^*(-\zeta)d\zeta\right)\right| \leq 1$ with equality if and only if $\psi(t)$ is an invariant wavelet.

*Proof*: The Cauchy inequality for complex signals establishes that

$$\left|\int_{-\infty}^{+\infty} F_1(x)F_2(x)dx\right|^2 \leq \int_{-\infty}^{+\infty} |F_1(x)|^2 \, dx \cdot \int_{-\infty}^{+\infty} |F_2(x)|^2 \, dx \quad (12)$$

with equality if and only if $F_1(x) = k F_2^*(x)$, where $k$ is an arbitrary constant. Therefore

$$\left|\frac{1}{\sqrt{2\pi}} \Re e\left(\int_{-\infty}^{+\infty} \psi(\zeta)\Psi^*(-\zeta)d\zeta\right)\right|^2 \leq$$

$$\frac{1}{2\pi} \int_{-\infty}^{+\infty} |\psi(\zeta)|^2 \, d\zeta \cdot \int_{-\infty}^{+\infty} |\Psi(\zeta)|^2 \, d\zeta = 1. \quad (13)$$

□

Thus, if $f \leftrightarrow F$ has a finite energy $E$, then

$$|<f, \frac{1}{\sqrt{2\pi}} F>| \leq E. \quad (14)$$

*Lemma 3*: (Gibbs inequality for wavelets) Given two continuous wavelets $\{\psi_i(t)\}_{i=1,2}$, then

$$-\int_{-\infty}^{+\infty} \psi_1^2(t) \log_2(\psi_1^2(t))dt \leq -\int_{-\infty}^{+\infty} \psi_1^2(t) \log_2(\psi_2^2(t))dt,$$

with equality if and only if $|\psi_1(t)| = |\psi_2(t)|$ ($\forall t$).

*Proof*: Applying the fundamental inequality $\ln x \leq x - 1$ with equality if and only if $x=1$ [6,14], one derives

$$\int_{-\infty}^{+\infty} \psi_1^2(t) \log_2\left(\frac{\psi_2^2(t)}{\psi_1^2(t)}\right) dt \leq$$

$$\frac{1}{\ln 2}\left(\int_{\{t|\psi_1(t) \neq 0\}} \psi_2^2(t)dt - \int_{-\infty}^{+\infty} \psi_1^2(t)dt\right) \leq 0.$$

(15)

□

A straightforward upper bound on the entropy of a wavelet can be derived.

*Proposition 4*: The global entropy of a wavelet $\psi(.) \leftrightarrow \Psi(.)$ is upper bounded by

Evoking the definition of the mutual information between two continuous random variables $X$ and $Y$, we recognise that the root for creating such a measure is defining some joint probability measure [16], [18]. We shall see that this can be done with the aide of the continuous wavelet transform (CWT), $CWT(a,b):=<f(t),\psi_{a,b}(t)>$. We infer that the energy conservation and the resolution of the identity [1] can be used so as to define a joint density between a signal and a wavelet in the *wavelet domain*.

*Definition 4*. (Joint signal-wavelet probability density). Let $f \in L^2(\mathbb{R})$ be a signal of energy $E := \int_{-\infty}^{+\infty} f^2(t)dt$. Given a mother continuous wavelet $\psi(.)$ with admissibility constant $c_\psi < +\infty$, the joint density between $f$ and a version $\psi_{a,b}(t)$, can be defined by

$$H_\psi \leq -\int_{-\infty}^{+\infty} \psi^2(\zeta) log_2\left(\frac{1}{2\pi}|\Psi(\zeta)|^2\right)d\zeta \qquad (16)$$
$$-\int_{-\infty}^{+\infty} \frac{1}{2\pi}|\Psi(\zeta)|^2 log_2\left(\psi^2(\zeta)\right)d\zeta,$$

with equality if and only if $\psi(.)$ is an invariant wavelet.
*Proof*: Follows from the lemma 3, with equality iff $\psi^2(\zeta) = \frac{1}{2\pi}|\Psi(\zeta)|^2$. □

Another interesting consequence of this approach is that the Kullback-Leiber distance [6,15] can assess the distance between two wavelets. Kullback called such a measure as the direct divergence, although alternative names have been suggested (e.g., cross-entropy, relative entropy). The notation $H(\psi_1 || \psi_2)$ by Cover and Thomas has thereafter been adopted [16].

*Definition 3*: (Wavelet distance). Given two continuous wavelets $\{\psi_i(t)\}_{i=1,2}$ compactly supported on $Supp(\psi_i(t))$, $i=1,2$ respectively, the Kullback discriminant can evaluate the distance between their two associated densities according to
$D_1(\psi_1, \psi_2) := H(\psi_1 || \psi_2) =$

$$2\int_{Supp(\psi_1)} \psi_1^2(t).log_2\left(\frac{\psi_1(t)}{\psi_2(\lambda_{1,2}.t)}\right)dt, \qquad (17)$$

where $\lambda_{1,2}$ is the ratio between the length of the support of the two wavelets, $\psi_1(t)$ and $\psi_2(t)$.

The distance $D_1(\psi, \psi_{Haar})$, for instance, plays the role of the divergence from equiprobability, $D_1$, introduced by Gatlin when investigating genetic messages [17].

From the practical viewpoint, this definition can be extended by considering the effective support of the wavelets instead of its (unbounded) support. The nonnegativity of $H(\psi_1 || \psi_2)$ is not the only reasoning for adapting it as a closeness measure between wavelets. Although being asymmetric and not holding the triangle inequality, its relevance relies on the Kullback principle. Moreover, a similar measure can be taken over the frequency domain. A distance measure between two wavelets, complete to a certain extent, could be

$$D_2(\psi_1, \psi_2) := 2\int_{effSupp(\psi_1)} \psi_1^2(t).log_2\left(\frac{\psi_1(t)}{\psi_2(\lambda_{1,2}.t)}\right)dt +$$
$$+ 2\int_{effSupp(\Psi_1)} |\Psi_1(w)|^2 .log_2\left(\frac{|\Psi_1(w)|}{|\Psi_2(\mu_{1,2}.w)|}\right)dw, \qquad (18)$$

where $\lambda_{1,2}$ (respectively $\mu_{1,2}$) is the ratio between the length of the effective support of the two wavelets in the time (respectively frequency) domain.

III. BY WAY OF A WAVELET INFORMATION THEORY

$$p(a,b) := \frac{|CWT(a,b)|^2}{E.c_\psi.|a|^2}, \quad a \neq 0. \qquad (19)$$

Clearly, $p(a,b)$ is a joint probability density (between the signal and the analysing wavelet) over the scale-translation domain. This joint density is thus expressed in terms of the *scalogram* of the signal, which visually provides information about local meaningful features embedded in the data [19]. Everything is now set for introducing the notion of wavelet mutual information to embroider on the WIT.

A. *Mutual Signal-wavelet Information*

*Definition 5:* (Mutual signal-wavelet information). The wavelet mutual information $I(f(t),\{\psi_{a,b}(t)\})$ between the signal and analysing wavelet can be found in a way somewhat similar to the classical definition $I(X;Y)=H(X)-H(X|Y)$ after using Bayes' rule:

$$I(f(t),\{\psi_{a,b}(t)\}) := \int_{-\infty}^{+\infty}\int_{-\infty}^{+\infty} \frac{|CWT(a,b)|^2}{E.c_\psi.|a|^2}.$$

$$log_2\left(\frac{E.c_\psi.|CWT(a,b)|^2}{\int_{-\infty}^{+\infty}|CWT(a,b)|^2 db \int_{-\infty}^{+\infty}|CWT(a,b)|^2 \frac{da}{a^2}}\right) dadb. \qquad (20)$$

This allows for computing the amount of information provided by the decomposition of a signal $f$ using a wavelet $\psi$.

Since $I(X,Y) \geq 0$ whatever the joint distribution $p(X,Y)$ [14,18], a nonnegative amount of information $I(f(t),\{\psi_{a,b}(t)\}) \geq 0$ is provided by the analysis.

The continuous time wavelet series (CTWS) can be used instead of the CWT. In the dyadic case, only $\left\{\psi_{n,m} = \frac{1}{\sqrt{2^m}}\psi(2^{-m}t - n)\right\}_{n,m\in Z}$ versions are considered.

Wavelet coefficients (details) are

$$w_{n,m} := <f,\psi_{n,m}> = \int_{-\infty}^{+\infty} f(t)\psi_{n,m}^*(t)dt, \; n,m\in Z, n\neq 0. \quad (21)$$

For orthogonal wavelets, the set $\{\psi_{n,m}(t)\}_{n,m\in Z}$, $t\in\mathbb{R}$, is a basis for $L^2(\mathbb{R})$ [1, Theorem 4.5, p.247] and therefore

$$f(t) = l.i.m. \sum_{n\in Z}\sum_{m\in Z} w_{n,m}\psi_{n,m}(t), \quad (22)$$

where *l.i.m.* denotes the limit in the mean and therefore

$$\sum_{n\in Z}\sum_{m\in Z} |w_{n,m}|^2 = \int_{-\infty}^{+\infty} f^2(t)dt = E. \quad (23)$$

The mutual signal-wavelet information can be derived from the homogeneous wavelet expansion as

$$I(f,\{\psi\}) = \sum_{n\in Z}\sum_{m\in Z} \frac{|w_{n,m}|^2}{E} log_2 \frac{E\cdot|w_{n,m}|^2}{\sum_{m'\in Z}|w_{n,m'}|^2 \sum_{n'\in Z}|w_{n',m}|^2}. \quad (24)$$

Finally, it is especially meaningful to evaluate the amount of information provided when the breathtakingly influential Mallat multiresolution approach is applied to the analysis of a signal [11,12]. Entropy can also be computed using detail coefficients (e.g., [20].)

*B. Multiresolution Information*

Suppose that $\{V_j\}$ constitutes the closed approximation subspaces of a multiresolution analysis (MRA) of $L^2(\mathbb{R})$ and let $\{W_j\}$ denote the sequence of orthogonal complementary (wavelet) spaces.

For any given level $j$, $clos(V_j \oplus W_j \oplus W_{j-1} \oplus ...) = L^2(\mathbb{R})$, where *clos* denotes the closure union and $\oplus$ indicates "orthogonal sums" of subspaces. An MRA expansion can be written in terms of scaling functions ($\phi$) and wavelet functions ($\psi$) as

$$f(t) = l.i.m. \sum_{k\in Z} v_{k,J}\phi_{k,J}(t) + \sum_{j=-\infty}^{J}\sum_{k\in Z} w_{k,j}\psi_{k,j}(t), \quad (25)$$

and

$$\sum_{k\in Z} |v_{k,J}|^2 + \sum_{j=1}^{J}\sum_{k\in Z} |w_{k,j}|^2 = \int_{-\infty}^{+\infty} f^2(t)dt = E, \quad (26)$$

where $v_J=\{v_{k,J}\}_{k\in Z}$ are the *approximation coefficients* and $w_j=\{w_{k,j}\}_{k\in Z}$, $j=1,2,...,J$ are the *detail coefficients* of the MRA. The mutual signal-wavelet information of an MRA can be computed from the inhomogeneous wavelet expansion by

$$I(f,\{\phi,\psi\}) = \sum_{k\in Z}\sum_{j=1}^{J} \frac{(|v_{k,J}|^2/J + |w_{k,j}|^2)}{E} \cdot$$

$$log_2 \frac{E\cdot(|v_{k,J}|^2/J + |w_{k,j}|^2)}{\left(\sum_{k'\in Z}(|v_{k',J}|^2 + |w_{k',j}|^2)\right)\left\{|v_{k,J}|^2 + \sum_{j'=1}^{J}|w_{k,j'}|^2\right\}} \quad (27)$$

IV. WAVELET-BASED INFORMATION PROCESSING OF A FEW SIGNALS

The main benefit of the MRA decomposition is the fact that there are plenty of room in the selection of the wavelet system. A specific signal feature can be enhanced in a particular decomposition level and different wavelets provide different readings. In order to gain some insight into the information theory approach for multiresolution analysis, a few simple discrete signals were analysed and their information content computed in each decomposition level. Three signals of length 16 were selected: the first one, shaped as a particular wavelet (db1); a dc-signal; and a random signal. Several orthogonal wavelet systems were considered: Daubechies (db1,…,db5), Coiflets (coif2, coif3), Symmlets (sym1, sym2).

Let $x_1[n]$=[9 11 9 11 9 11 9 11 9 11 9 11 9 11 9 11] be the first testing signal, which is a version of the Haar wavelet (db1). MRA-info for this signal is shown in Tables I, II and III.

TABLE I.
INFORMATION AMOUNT IN THE ONE-LEVEL MRA-DECOMPOSITION OF THE SIGNAL $x_1[n]$.

| wave | Approximation (%) | Detail 1 (%) | Total infor. amount (100%) |
|---|---|---|---|
| db1 | 0.0000 (0.0) | <0.0001 (100) | <0.0001 |
| db2 | <0.0001 (0.9) | 0.0001 (99.1) | 0.0001 |
| db4 | <0.0001 (0.7) | 0.0018 (99.3) | 0.0018 |
| db5 | <0.0001 (0.6) | 0.0027 (99.4) | 0.0028 |
| coif2 | <0.0001 (0.8) | 0.0007 (99.2) | 0.0008 |
| coif3 | <0.0001 (0.8) | 0.0007 (99.2) | 0.0007 |
| sym1 | 0.0001 (26.6) | 0.0002 (73.4) | 0.0003 |
| sym2 | 0.0001 (26.6) | 0.0003 (73.4) | 0.0004 |

TABLE II.
INFORMATION AMOUNT IN THE TWO-LEVEL MRA-DECOMPOSITION OF THE SIGNAL $x_1[n]$.

| wave | Approx. (%) | Detail 2 (%) | Detail 1 (%) | Total infor. amount (100%) |
|---|---|---|---|---|
| db1 | 0.0071 (20.2) | 0.0000 (0.0) | 0.0280 (79.8) | 0.0351 |
| db2 | 0.0043 (20.3) | 0.0001 (0.7) | 0.0169 (79.0) | 0.0214 |
| db4 | 0.0030 (17.5) | 0.0008 (4.9) | 0.0135 (77.6) | 0.0174 |
| db5 | 0.0022 (16.2) | 0.0004 (2.6) | 0.0111 (81.2) | 0.0136 |
| coif2 | 0.0011 (17.4) | 0.0004 (6.9) | 0.0048 (75.7) | 0.0063 |
| coif3 | 0.0000 (0.6) | 0.0003 (35.7) | 0.0006 (63.7) | 0.0010 |
| sym1 | 0.0071 (20.0) | 0.0002 (0.6) | 0.0283 (79.4) | 0.0356 |
| sym2 | 0.0044 (20.1) | 0.0003 (1.6) | 0.0171 (78.3) | 0.0218 |

TABLE III.
INFORMATION AMOUNT IN THE THREE-LEVEL
MRA-DECOMPOSITION OF THE SIGNAL $x_1[n]$.

| wave | Approx. (%) | Detail 3 (%) | Detail 2 (%) | Detail 1 (%) | Total infor. amount (100%) |
|---|---|---|---|---|---|
| db1 | 0.0106 (19.3) | 0.0000 (0.0) | 0.0000 (0.0) | 0.0445 (80.7) | 0.0552 |
| db2 | 0.0076 (19.7) | 0.0001 (0.2) | 0.0005 (1.2) | 0.0306 (78.9) | 0.0388 |
| db4 | 0.0032 (18.4) | 0.0001 (0.4) | 0.0007 (4.2) | 0.0133 (77.0) | 0.0172 |
| db5 | 0.0042 (17.9) | 0.0001 (0.6) | 0.0003 (1.3) | 0.0189 (80.2) | 0.0235 |
| coif2 | 0.0021 (18.7) | 0.0001 (0.7) | 0.0004 (3.5) | 0.0087 (77.1) | 0.0113 |
| coif3 | 0.0000 (0.5) | 0.0001 (7.5) | 0.0003 (31.1) | 0.0007 (60.9) | 0.0011 |
| sym1 | 0.0107 (19.1) | 0.0001 (0.1) | 0.0002 (0.4) | 0.0448 (80.4) | 0.0557 |
| sym2 | 0.0077 (19.5) | 0.0001 (0.3) | 0.0007 (1.7) | 0.0308 (78.5) | 0.0393 |

For the dc signal, null information content was found for each and every one approximation, whatever the decomposition level, as expected. As another naïve example, the following random signal of length 16 was generated:

$x_3[n]$=[0.9501 0.2311 0.6068 0.4860 0.8913 0.7621 0.4565 0.0185 0.8214 0.4447 0.6154 0.7919 0.9218 0.7382 0.1763 0.4057].

TABLE IV.
INFORMATION AMOUNT IN THE ONE-LEVEL
MRA-DECOMPOSITION OF THE SIGNAL $x_3[n]$.

| wave | Approximation (%) | Detail 1 (%) | Total Infor. amount (100%) |
|---|---|---|---|
| db1 | 0.0111 (10.5) | 0.0944 (89.5) | 0.1055 |
| db2 | 0.0112 (13.3) | 0.0730 (86.7) | 0.0842 |
| db4 | 0.0088 (9.4) | 0.0846 (90.6) | 0.0934 |
| db5 | 0.0153 (12.4) | 0.1081 (87.6) | 0.1234 |
| coif2 | 0.0075 (9.6) | 0.0705 (90.4) | 0.0780 |
| coif3 | 0.0090 (10.7) | 0.0752 (89.3) | 0.0842 |
| sym1 | 0.0210 (15.1) | 0.1182 (84.9) | 0.1392 |
| sym2 | 0.0183 (15.9) | 0.0969 (84.1) | 0.1152 |

TABLE V.
INFORMATION AMOUNT IN THE TWO-LEVEL
MRA-DECOMPOSITION OF THE SIGNAL $x_3[n]$.

| wave | Approx. (%) | Detail 2 (%) | Detail 1 (%) | Total Infor. amount (100%) |
|---|---|---|---|---|
| db1 | 0.0312 (13.4) | 0.0982 (42.3) | 0.1028 (44.3) | 0.2322 |
| db2 | 0.0282 (22.6) | 0.0161 (12.9) | 0.0806 (64.5) | 0.1249 |
| db4 | 0.0233 (15.9) | 0.0357 (24.3) | 0.0881 (59.8) | 0.1471 |
| db5 | 0.0144 (8.9) | 0.0378 (23.4) | 0.1093 (67.7) | 0.1615 |
| coif2 | 0.0055 (5.0) | 0.0385 (35.3) | 0.0651 (59.7) | 0.1091 |
| coif3 | 0.0168 (11.9) | 0.0342 (24.3) | 0.0897 (63.6) | 0.1407 |
| sym1 | 0.0625 (18.8) | 0.1109 (33.3) | 0.1594 (47.9) | 0.3328 |
| sym2 | 0.0582 (27.3) | 0.0175 (8.2) | 0.1373 (64.5) | 0.2131 |

TABLE VI.
INFORMATION AMOUNT IN THE THREE-LEVEL
MRA-DECOMPOSITION OF THE SIGNAL $x_3[n]$.

| wave | Approx. (%) | Detail 3 (%) | Detail 2 (%) | Detail 1 (%) | Total Infor. amount (100%) |
|---|---|---|---|---|---|
| db1 | 0.0991 (28.8) | 0.0020 (0.6) | 0.1324 (38.6) | 0.1098 (32.0) | 0.3433 |
| db2 | 0.0656 (25.2) | 0.0241 (9.3) | 0.03 (11.6) | 0.1401 (53.9) | 0.2598 |
| db4 | 0.0301 (17.6) | 0.0111 (6.5) | 0.0424 (24.9) | 0.087 (51.0) | 0.1706 |
| db5 | 0.0233 (12.0) | 0.0318 (16.2) | 0.0333 (17.1) | 0.1067 (54.7) | 0.1952 |
| coif2 | 0.0107 (8.2) | 0.0202 (15.4) | 0.0454 (34.8) | 0.0544 (41.6) | 0.1307 |
| coif3 | 0.0077 (5.3) | 0.039 (26.9) | 0.0368 (25.4) | 0.0613 (42.4) | 0.1448 |
| sym1 | 0.1051 (24.1) | 0.0439 (10.1) | 0.149 (34.2) | 0.1374 (31.6) | 0.4353 |
| sym2 | 0.073 (21.5) | 0.0656 (19.3) | 0.0329 (9.7) | 0.1678 (49.5) | 0.3394 |

The computation of the information content was carried out using the Matlab[TM]. MRA-info of $x_3[n]$ is presented in Tables IV, V, and VI. Overall, sym1-based MRA provides much information and coif2-based MRA provides less information for this particular signal.

V. CONCLUSIONS

This paper intends to present a new regard to wavelets, which can be viewed as an information processing technique. Defining the mutual information between signals and analysing wavelets then fulfilled the heart of a wavelet information theory. This approach may help selecting suitable wavelets for analysing a signal. Distances between two wavelets were proposed and it was shown that $D_1(\psi, \psi_{Haar})$ is a divergence measure from equiprobability. The emphasis of this paper was on conveying the chief ideas behind the WIT as opposed to presenting a formal mathematical development. Despite the fact that a mere draft of this technique has been outlined, it instigates expectations on both theoretical and practical information theory outcomes related to the wavelet analysis (e.g. criteria in the choice of the wavelet system; waveshrink, wavelet compression carried out on wavelet-based information-theory-oriented algorithms not on the energy.)

ACKNOWLEDGMENTS

The first author is obliged to Professor Valdemar C. da Rocha Jr, who was steadfast in creating an exceptional atmosphere of enthusiastic, inspiring interest in IT.